\documentclass[%
 reprint,
 amsmath,amssymb,
 aps,
floatfix,
]{revtex4-2}

\usepackage{graphicx}
\usepackage{natbib}
\usepackage{dcolumn}
\usepackage{bm}
\usepackage{booktabs}
\usepackage{xcolor}

\usepackage{siunitx}
\usepackage{stfloats}

\usepackage{hyperref}

\usepackage{import}
\usepackage{transparent}
\usepackage{float}

\usepackage{xfp}
\usepackage{adjustbox}
\newlength\myHeight 
\newlength\myWidth
\NewDocumentCommand{\doCalc}{m}{$
    \fpeval{#1}
 $}
 
\newcommand{\wordsSingleCol}{\\\textbf{Words for \Figure: \doCalc{ceil(150 / (\the\myWidth / \the\myHeight) + 20)}}}
\newcommand{\wordsDoubleCol}{\\\textbf{Words for \Figure: \doCalc{ceil(300 / (0.5* \the\myWidth / \the\myHeight) + 40)}}}

\newcommand{\Eq}[1]{Eq. #1}
\newcommand{\Sec}[1]{Section #1}

\newcommand{\Figure}{Fig.~}

\newcommand{\ot}{\hat{t}}
\newcommand{\state}{y}
\newcommand{\istate}{\state_0}
\newcommand{\augstate}{s}
\newcommand{\eom}{f}
\newcommand{\params}{p}
\newcommand{\loss}{L}
\newcommand{\ad}{a}
\newcommand{\grad}{g}
\newcommand{\jac}[2]{\mathcal{J}_{#1,#2}}

\newcommand{\eomof}{f(\state,t,\params)}
\newcommand{\jacof}[2]{\mathcal{J}_{#1,#2}(\state, t, \params)}

\newcommand{\tder}[1]{\frac{d}{dt}#1}
\newcommand{\momder}[2]{\frac{\partial #2}{\partial #1}}
\newcommand{\conder}[2]{\frac{\mathcal{D} #2}{\mathcal{D} #1}}

\newcommand{\at}[1]{_{(#1)}}
\begin{document}

\title{AdoptODE: Fusion of data and expert knowledge for modeling dynamical systems}

\author{Leon Lettermann}
\altaffiliation[Now at: ]{BioQuant \& Institute for Theoretical Physics, Heidelberg University, 69120 Heidelberg, Germany}
 \author{Alejandro Jurado}
 \author{Timo Betz}
 \author{Florentin  W\"org\"otter}
  \altaffiliation[Also: ]{Bernstein Center for Computational Neuroscience, Friedrich-Hund Platz 1, 37077 G\"ottingen, Germany}
\author{Sebastian Herzog}%
 \email{sherzog3@gwdg.de}
\affiliation{%
 Third Institute of Physics - Biophysics, Georg-August Universi\"at G\"ottingen, 37077 G\"ottingen, Germany\\
}%

\date{\today}

\begin{abstract}
Building a representative model of a complex system remains a highly challenging problem. While by now there is basic understanding of most physical domains, model design is often hindered by lack of detail, for example concerning model dimensions or its relevant constraints. Here we present a novel model-building approach -- adoptODE -- augmenting basic system descriptions, based on expert knowledge in the form of ordinary differential equations, with continuous adjoint sensitivity analysis related to artificial neural network principles, based on observable data. With this we have created a general tool, that can be applied to any physical system described by ordinary differential equations. AdoptODE allows validating or extending the initial description, for example with different variables and constraints. This way one arrives at a better-optimised, representative low-dimensional model, which can fit existing data and predict novel experimental outcomes. 

\end{abstract}

\maketitle


In recent years, artificial neural networks (ANNs) have led to significant breakthroughs in many areas of science (e.g.,protein folding \cite{alphafold2}, three-body problem \cite{threebody}, material properties \cite{TianEtAl2018}, etc.).
Most of these are 'black box' models, where predictions are made by indirect parameterization of the data via the weights of the ANN, which cannot be directly interpreted. 
By contrast, approaches where data is automatically parameterised as variables belonging to a set of equations,  would lead to a better-understandable model of the underlying system.

\begin{figure*}[ht]
    \center
    \adjustbox{gstore width=\myWidth, gstore height=\myHeight, center}{
    \includegraphics{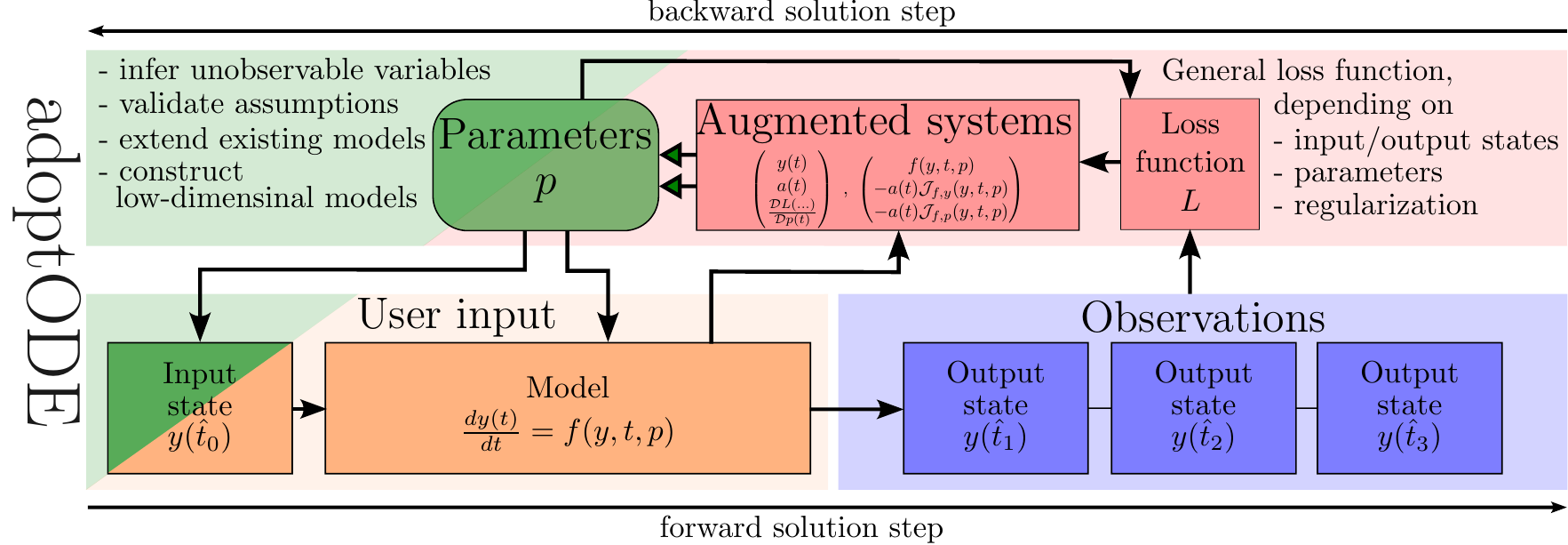}
    }
    \caption{Diagram of adoptODE. A prior-provided ODE model generates outputs depending on its parameters and the input state. By comparing to data and possible additional constraints formulated in the loss function $L$, adoptODE optimizes input states and parameters. The augmented system (defined in Sec.~\ref{sec:Augmented}) required for the optimisation of the parameters is set up and solved automatically by adoptODE. The user input is coloured in orange and the data used to find the desired parameters is coloured in blue.}
    \label{fig:method}
\end{figure*}

To achieve this, we present -- adoptODE -- a modeling approach that uses data to infer \textit{explicit} representations applicable in a wide range of problems, relying on the fact that for most (non-linear) systems there is to-date at least some fundamental knowledge about the underlying physics. Accordingly, we use this to guide the parameterization process by providing our inference-process up-front with representative ordinary differential equations (ODEs), where some or all of the parameters representing unknown quantities of the ODEs are optimised to fit the data. Here we rely on the automatic differentiation power of modern deep learning frameworks like JAX \cite{jax2018github}, which we utilise to formulate a method for continuous, adjoint sensitivity analysis (CASA). This allows the efficient optimisation of parameter-dependent ordinary differential equations with thousands of parameters to be efficiently optimised on conventional graphics processing units. Convergence is assured by utilising a modern, powerful optimisation algorithm (e.g., \cite{tjoa1991simultaneous, farmer1991optimal, baake1992fitting, tanartkit1995stable, li2005parameter}, for a review see \cite{brewer2008fitting}). In this way, we arrive at an interpretable system representation that achieves one or more of the following (1) infer unobservable variables from observable data, (2) validate model assumptions, (3) extend existing models with new variables to better fit the data and (4) construct low-dimensional predictive models.
AdoptODE takes inspiration from the concept of \textit{Neural} Ordinary Differential Equations (NODE), \cite{chen2018neural}, where the central idea is that a sequence of residual blocks \cite{resnet} in an ANN represents a solution of the ODE $\frac{dz(t)}{dt}=f(z,t, p)$, integrated by the Euler method. In such residual networks an initial state $z_{(t=0)}=z_0$ is propagated and at each layer modified according to the response of that layer  $f_k$ as ${z_{k}=z_{k-1}+f_k(z_{k-1})}$. The networks depth $k$ corresponds to the time $t$ and the equation of motion depends on parameters $p$. For the classical NODE approach \cite{chen2018neural}, the equation of motion (EOM) is taken as a fully connected ANN. Instead of using a black box model, like an ANN, an expert guess for the EOM is utilised in adoptODE for predicting the system's dynamics based on incomplete data. AdoptODE can now obtain the parameters $p$ in the EOMs $f(z,t,p)$, which are meaningful and interpretable quantities. For adoptODE only the data and the system definition, including the ODE (hence, the EOM), have to be provided, but no changes of the optimisation routine are required (\Figure\ref{fig:method}).

To demonstrate performance and the wide range of problems on which this can operate, our approach is evaluated on five quite different problem classes. 1. Kolmogorov model: Lotka Volterra model \cite{lotka2002contribution,volterra1926fluctuations}, 2. Particle model: Interactive N-body systems \cite{newton1833philosophiae}, 3. Excitable media: Bueno-Orovio–Cherry–Fenton model \cite{bueno2008minimal}, 4. Fluid dynamics: Rayleigh-B\'enard Convection \cite{rayleigh1916lix}, 5. Real experimental data: Zebrafish embryogenesis.  We emphasise that the code, used for this, is rather simple and can easily be adapted to address other, specific problems, too, with the potential to create wider impact on scientific model building (The complete framework, called adoptODE, can be found at: \url{https://gitlab.gwdg.de/sherzog3/adoptode.git})

\section{\label{sec:results}Results}

We validate adoptODE on five, different problem domains. (1) Kolmogorov model: Lotka Volterra model where we show the application of adoptODE to continuous-time Markov processes and the performance of adoptODE when working with noisy data and wrong model assumptions. (2) Particle model: Interactive N-body, is a demonstration of the scalability of adoptODE and that even interactive system with a high number of elements can be reconstructed with high precision. (3) Excitable media (heart dynamics) by the Bueno-Orovio–Cherry–Fenton model: AdoptODE can reconstruct the parameters of a high-dimensional model and fields for diffusion driven system for chaotic behaviour. (4) Fluid dynamics: Rayleigh-B\'enard Convection where a complete unknown field, the temperature, can be extracted from only velocity data. (5)  New experimental data of Zebrafish embryogenesis: This is a case where we extend existing models with new variables by which novel unexpected biophysical results could be obtained. We define for all scenarios the \textit{initial state}, the assumed EOM with its \textit{parameters}, the \textit{input data} and we ask the system to solve a specific \textit{task}. 

\subsection{\label{sec:results_lv}Lotka-Volterra model}

\begin{figure}[h]
    \center
    \includegraphics{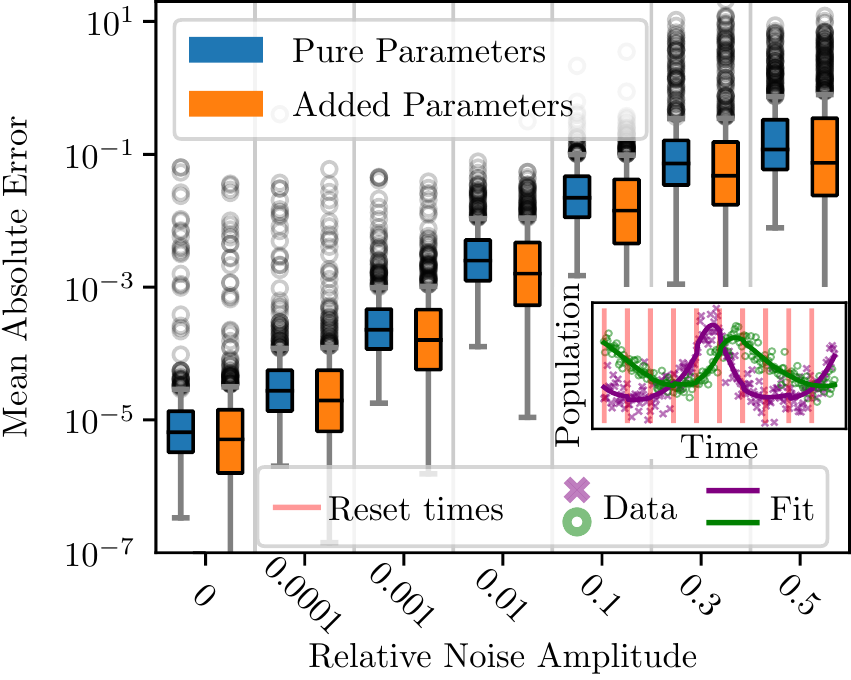}
    \caption{Results for the Lotka-Volterra model, showing statistics over 500 randomly generated systems for each noise level. The absolute errors were averaged over the four parameters of the pure model and the two added parameters separately. For the added parameters, the MAE is the mean absolute value as the ground truth is always zero. The inset shows and example trajectory with reset times, at which the system  is reset to new, simultaneously trained initial conditions. The error of the initial conditions behaves comparably to the error of the parameters (not displayed).}
    \label{fig:lv}
\end{figure}
The Lotka-Volterra model\cite{lotka2002contribution,volterra1926fluctuations} represents a system of two first order non-linear, coupled differential equations. It shall serve as an example for the Kolmogorov model class to characterise continuous-time Markov processes. 

\noindent\underline{State}: Current population of prey ($x$) and predators ($y$).

\noindent\underline{EOM}: 
\begin{equation*}
    \dfrac{\mathrm{d}x}{\mathrm{d}t}= \alpha x - \beta x y \,\textcolor{orange}{\Big[} +c_1x^2\textcolor{orange}{\Big]},\quad \dfrac{\mathrm{d}y}{\mathrm{d}t}=-\gamma y + \delta x y\, \textcolor{orange}{\Big[}+ c_2y^2\textcolor{orange}{\Big]},
\end{equation*}
where the $c_i$ terms are additional terms, added to the pure model in recovery to demonstrate that adoptODE can be used to identify correct terms in model development.

\noindent\underline{Parameters}: $\alpha, \beta, \gamma, \delta \in \mathbb{R}^+$ and $c_1,c_2 \in \mathbb{R}$. 

\noindent\underline{Input data:} 500 trajectories with random parameter values and initial populations drawn from $[0.1,1]$ were generated at each noise level with the pure model ($c_1=c_2=0$). Each observed datapoint was perturbed by random noise with amplitude proportional to its value and the given relative noise amplitude. 

\noindent\underline{Task}: Infer model parameters, here $\alpha, \beta, \gamma, \delta$, and the initial conditions for $x,y$ from observed, noisy trajectories. The added terms should be quantified as insignificant for the system by finding close to zero values for the parameters $c_1,c_2$

\noindent\underline{Results}: The mean absolute errors of parameter recovery are on average less than 0.1 for all noise amplitudes (\Figure\ref{fig:lv}),  indicating that parameter recovery was successful. At large noise levels the error sometimes exceeded $1$, signifying parameters out of range. In practice, such noise levels, will, however, likely not play any role. 
The values for the added parameters $c_1,c_2$ are similar to the errors of the pure parameters, meaning they are orders of magnitude smaller than the pure parameters for lower noise ratios, allowing to reject the respective terms as irrelevant for the model.

\subsection{\label{sec:results_nb}Interactive N-Body problem}
The N-body problem shall be considered as an example of a dynamical system consisting of particles, which influence each other by physical forces. Here we consider the case where $N$ point masses interact via some pairwise force $F$. 

\noindent\underline{State}: Positions and velocities of all bodies.

\noindent\underline{EOM}: 
\begin{equation*}
     \dfrac{\mathrm{d}^2x}{\mathrm{d}t^2}= \frac{1}{m_i}\sum_j F (x_i - x_j).
\end{equation*}

\noindent\underline{Sub-problem 1: Gravitational Systems:}
The first problem is based on Newton's law of gravity, which is used to describe the orbits of the planets in the solar system.

\noindent\underline{Parameters}: Masses of the planets up to Saturn and of the sun.  

\noindent\underline{Input data:} From NASA's Horizon Systems \cite{giorgini2015status} the trajectories of the sun, the planets up to Saturn and as additional force probes the asteroids Apophis and Eros, sampled every 20 days throughout the 20th century.

\noindent\underline{Task}: Infer model parameters, here mass, from trajectories. 

\begin{figure}[ht]
    \center
    \adjustbox{gstore width=\myWidth, gstore height=\myHeight, center}{
    \includegraphics{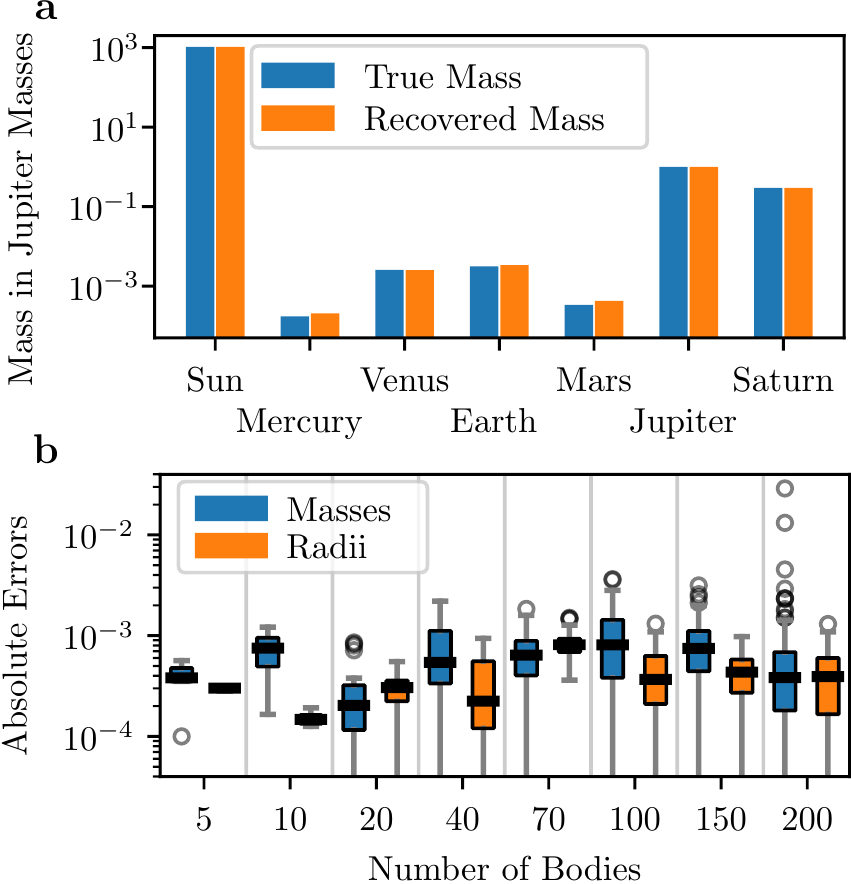}
    }
    \caption{Results for a: the solar system up to Saturn as an example of attractive N-body interactions and b: the repulsive  N-body dynamics. Distributions are the errors of different bodies in one simulation.}
    \label{fig:nbody2}
\end{figure}

\noindent\underline{Results}: Although the solar system is dominated by the sun's mass, so that the planetary masses can only be inferred from the perturbations they cause, recovery is successful as shown in \Figure a: \ref{fig:nbody2}, with the largest deviation for Mars at a relative error of 0.3 and less than 1\% error for Saturn, Jupiter and the sun.\\

\noindent\underline{Sub-problem 2: Repulsive systems:} The second example demonstrates repulsive interactions of spherical particles with different effective radii, described by the Lennard-Jones potential \cite{jones1924determination}.

\noindent\underline{Parameters}: Mass and radius of each particle. \underline{Input data:} For every number of bodies, random radii and masses were drawn from the intervals $[0.2,0.7]$ and $[1,10]$ respectively. The spheres were randomly assigned positions on a grid (to avoid initial overlaps in high density runs) with initial velocity drawn from a truncated normal distribution over $[-2,2]$.

\noindent\underline{Task}: As above but also: infer non-observable variables from observable data. Here masses and radii are obtained from the trajectories, where (different from the gravitational interaction, above) this information is only contained in the collision events.

\noindent\underline{Results}: In \Figure\ref{fig:nbody2} it can be seen that for all but two masses at the largest system the recovery of masses and radii was within 1\% accuracy, taking the smallest value for masses and radii as reference.

\subsection{\label{sec:results_bocf}Bueno-Orovio–Cherry–Fenton Model}
The Bueno-Orovio–Cherry–Fenton Model is a minimal ionic model for human ventricular cells in the heart. It shall be considered as an example for excitable media. In general an excitable medium is a nonlinear dynamical system to describe the propagation of some waves, which can influence each other, but where one wave cannot propagate through the other. Such systems are used for modelling of many bio-chemical domains. 

\noindent\underline{State}: Four fields (voltage and three gating variables), each sampled at a $512 \times 512$ square lattice.

\noindent\underline{EOM}: The (cumbersome) equations of the BOCF model~\cite{bueno2008minimal}.

\noindent\underline{Parameters}: The parameters of the BOCF model influence each other, so different sets of parameters can cause the same dynamics of the model. To ensure comparability between the true parameters and the recovered parameters, we first identified the 10 most influential parameters by separately perturbing them by a factor of 1.5 and comparing the resulting mean absolute errors. The 10 parameters out of the 28 parameters with the highest resulting mean absolute error were then selected for the recovery task. These 10 parameters were then perturbed by random factors between $1/2$ and $3/2$ and recovered by adoptODE.

\noindent\underline{Input data}: A procedure of repeated excitation generates a random initial state showing spiral-wave patterns. From this initial state, a trajectory with 100 time points over \SI{50}{\milli\second} is generated.

\noindent\underline{Task}: (1) infer non-observable variables from observable data, (2) validate model assumptions. Here: Find parameter values to best describe a given trajectory.

\begin{figure*}[b]
    \center
    \adjustbox{gstore width=\myWidth, gstore height=\myHeight, center}{
    \includegraphics{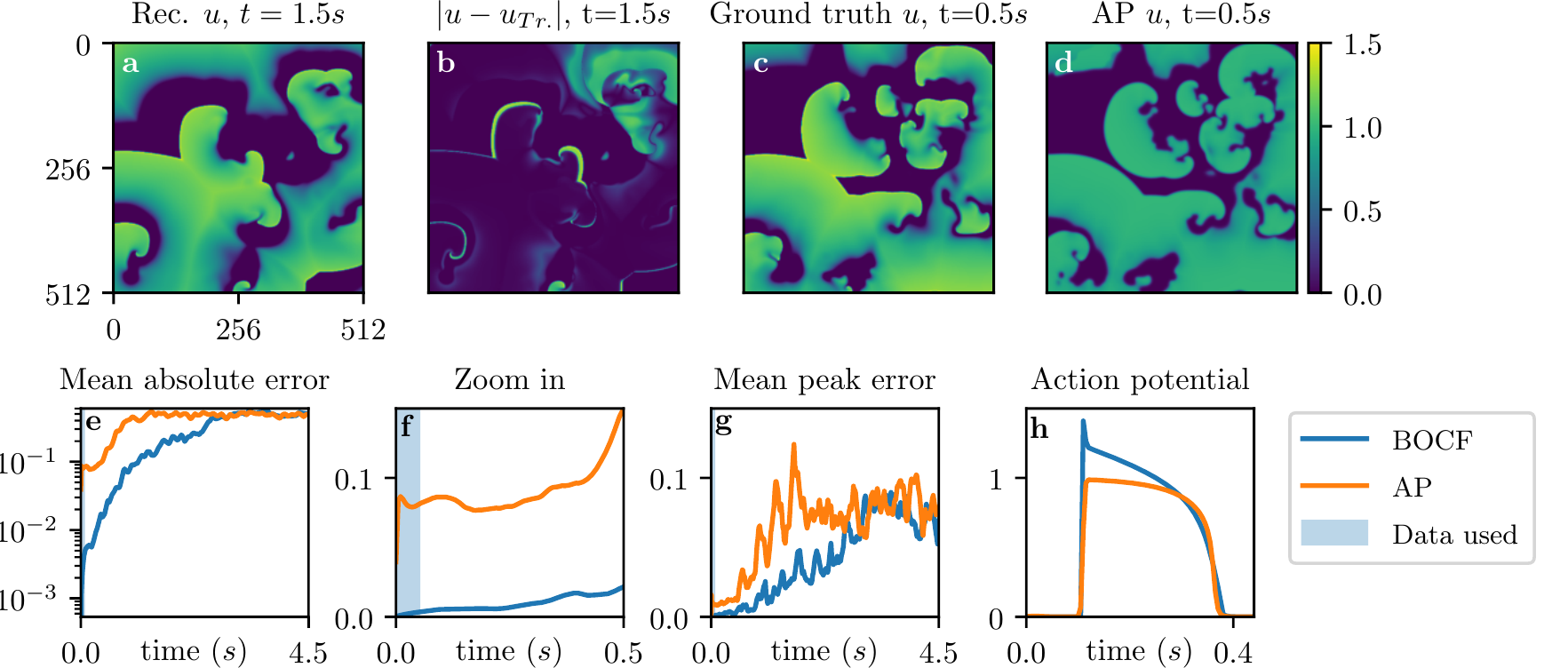}
    }
    \caption{Results of BOCF model. a-d: Excitation field $u$, value given by colour bar on the right. a: Recovered BOCF field after \SI{1.5}{\second}. b: Difference between recovery in a and target. c: Truth after \SI{0.5}{\second}. d: Recovery using the Aliev-Panfilov model. e: Mean absolute error of the recovery using the BOCF model and the AP model on a logarithmic scale. f: Zoom in of e, the length of the input sequence used to optimise the parameters via adoptODE is shown in blue. Also shown in e and g, but difficult to see because of the short sequence. g: Error in peak occurrence time averaged over all pixels. h: Shape of the first action potential at the center pixel. See supplement video V1.}
    \label{fig:bocf}
\end{figure*} 

Additionally, we show how to use adoptODE to reproduce the BOCF model excitation field using the simpler Aliev-Panfilov (AP) model \cite{aliev1996simple}, a model with only six free parameters and two fields. Because the form of the excation front and the single gating field in the AP model are different from the BOCF model and its three gating fields, both initial fields have to be reconstructed simultaneously with the parameters, adding $\approx \num{5e5}$ degrees of freedom to the optimization. Taking this into account, adoptODE finds parameters for which AP reproduces the BOCF dynamics, even if the prediction breaks down faster than the BOCF recovery.

\noindent\underline{Results}: To demonstrate the versatility of the recovery, for the BOCF system we analyse the growing discrepancy between the true and the recovered system with time, where the recovery was once calculated using the BOCF model, and again with the AP model where additionally the initial fields had to be recovered. Note that for the investigated parameters the BOCF model exhibits chaotic behaviour (maximum Lyapunov exponent $\lambda_{\text{max}}\approx 0.00276$), i.e. even small deviations lead to an exponentially increasing error. For the BOCF model, only a very short time (\SI{50}{\milli\second} (\Figure\ref{fig:bocf}e-g blue area)) had to be observed for a  meaningful recovery. Measured by the mean absolute error (\Figure\ref{fig:bocf}e on a logarithmic scale and \Figure\ref{fig:bocf}f a zoom-in of the initial segment on a linear scale) and the mean error of peak occurrence averaged over all pixels (\Figure\ref{fig:bocf}g), recovery was largely valid for up to 2 seconds. The errors at 1.5 seconds were mostly small differences in the propagation of individual excitation fronts and only in the upper right area already larger deviations (\Figure\ref{fig:bocf}a and b) were found. For the AP model, \SI{0.5}{\second} were used for training. Over this period the recovery is surprisingly good,  considering  the AP model cannot truly capture the BOCF dynamic as visible from the different shape of action potentials (\Figure\ref{fig:bocf}g). Here, the optimisation probably uses the fit of the initial conditions for the gating field to compensate for the different models, leading to a quick divergence after the observed time (cf. Supplementary Figure S1 displaying the resulting initial fields for the AP model). However, if we compare the ground truth $u$ field at $t=0.5s$ from the BOCF model (\Figure\ref{fig:bocf}c) with the recover $u$ field for the same time (\Figure\ref{fig:bocf}d), then we see that the main features are well represented by the AP model. The recovered parameters for the BOCF model had a mean absolute deviation of $0.76 \%$ and a median absolute deviation of $0.4 \%$, the recovered values are shown in supplementary table 1.

\subsection{\label{sec:results_rbc}Rayleigh-B\'enard Convection}
Rayleigh-B\'enard convection is an example of self-organising nonlinear chaotic systems in the field of fluid thermodynamics. A fluid enclosed in a volume, the bottom of which is heated, creates convective flow patterns conveying heat from the lower to the upper plate. Such convection cells are an example for the formation of dissipative structures far from thermodynamic equilibrium. The structure of these convection patterns is governed by the dimensionless Rayleigh and Prandtl numbers.

\noindent\underline{State}: Temperature perturbation $T$ and 2D velocity field $\vec u$, sampled at a $100 \times 100$ square lattice with periodic boundary conditions from right to left.

\noindent\underline{EOM}: Equations for Rayleigh-B\'enard Convection in the Oberbeck-Boussinesq approximation \cite{oberbeck1879warmeleitung, boussinesq1903thorie}, including a prescription to obtain the pressure field $p$ necessary in every step to maintain an incompressible flow \cite{rempfer2006boundary}.

\noindent\underline{Parameters}: Rayleight and Prandtl numbers are assumed to be known, initial temperature field is unknown. 

\noindent\underline{Input data}: Trajectories obtained from simulation at Prandtl-Number $7.0$ and Rayleigh-Number $10^7$, but only the velocity field is available to the system for training, as would be the case in data from Particle Image Velocimetry (PIV, \cite{grant1997particle}) experiments. Measured in Lyapunov-times $\tau$ estimated from the same data, $0.35\tau$ was used for training.

\noindent\underline{Task}: (1) infer non-observable variables from observable data, (2) validate model assumptions. Here: Reconstruct the temperature field from the velocity field.

\begin{figure}[h]
    \center
    \adjustbox{gstore width=\myWidth, gstore height=\myHeight, center}{
    \includegraphics{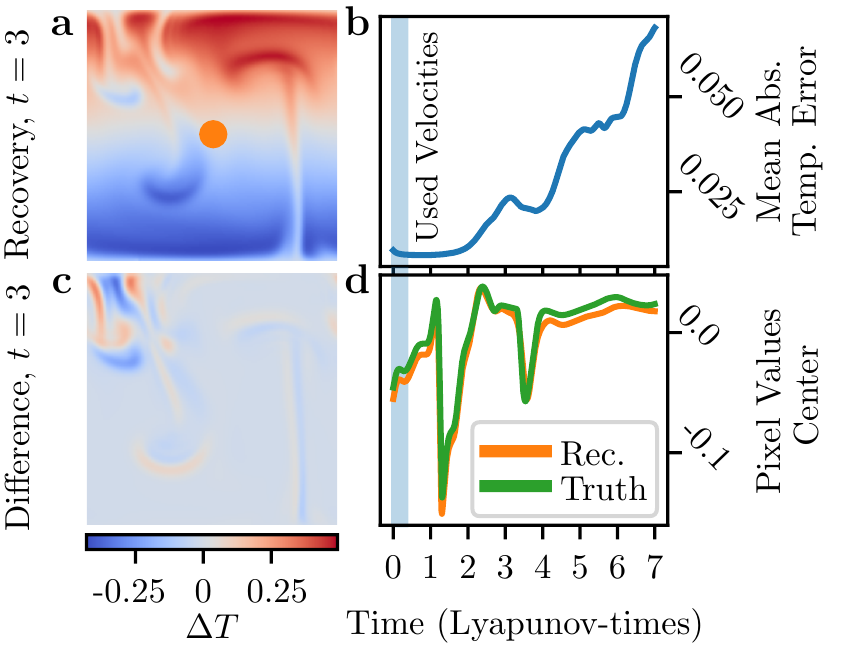}
    }
    \caption{Results for 2D Rayleigh-B\'enard Convection. a: Temperature perturbation. b: Mean absolute error between recovered and true temperature perturbation. The length of the input sequence used to optimise the parameters via adoptODE is shown in blue c: Difference from ground truth. d: Pixelvalue of centre pixel.}
    \label{fig:RB}
\end{figure}

\noindent\underline{Results}: Similarly to the BOCF Model, we validate the practicality of the recovered temperature field (\Figure\ref{fig:RB}a) by its power to predict the system longer than observed for training. The difference between the ground truth data and the recovered temperature field is shown in (\Figure\ref{fig:RB}c). We find the recovery to be accurate for around \si{2}$\tau$ (\Figure\ref{fig:RB}b), and still qualitatively correct at \si{3}$\tau$ (\Figure\ref{fig:RB}b and d), before it begins to more strongly deviate from the truth at \si{7}$\tau$. This behaviour is best seen in supplementary video V2.

\subsection{\label{sec:results_ze}Zebrafish Embryogenesis}
The case of zebrafish embryogenesis shows the use of adoptODE on real experimental data. Zebrafish embryos, genetically modified to show fluorescent nuclei, were imaged in a light-sheet microscope from 5 to 7.5 hours after fertilisation (hpf). The lightsheet scans the whole embryo once every 180 seconds and generates a full 3D image from which the position of every cell can be extracted. Afterwards, cells can be tracked from one frame to the next using Gaussian mixture models implemented in the library TGMM \cite{amat2014fast}, resulting in branching trees of trajectories, as cells do actively divide during the time observed. The trajectories obtained in this way are not perfect and contain noise due to the experimental setup. To apply a continuous hydro-dynamical flow model supplemented with active stresses, from the trajectories a mean velocity field was approximated by binning differences in cell positions between frames. The embryo develops on one pole of its spherical yolk cell. Around 5~hpf this cap spreads around the yolk cell in a process called epiboly, which coincides with and is required for the development of the germ layers \cite{kimmel1995stages}. To study the influence of active processes on this spreading, a simplified two-dimensional Navier-Stokes flow on a sphere is assumed for the embryo on-top of the yolk, and supplemented by a 2D time-dependent vector field of active forces. 

\begin{figure*}[h]
    \centering
    \adjustbox{gstore width=\myWidth, gstore height=\myHeight, center}{
    \includegraphics{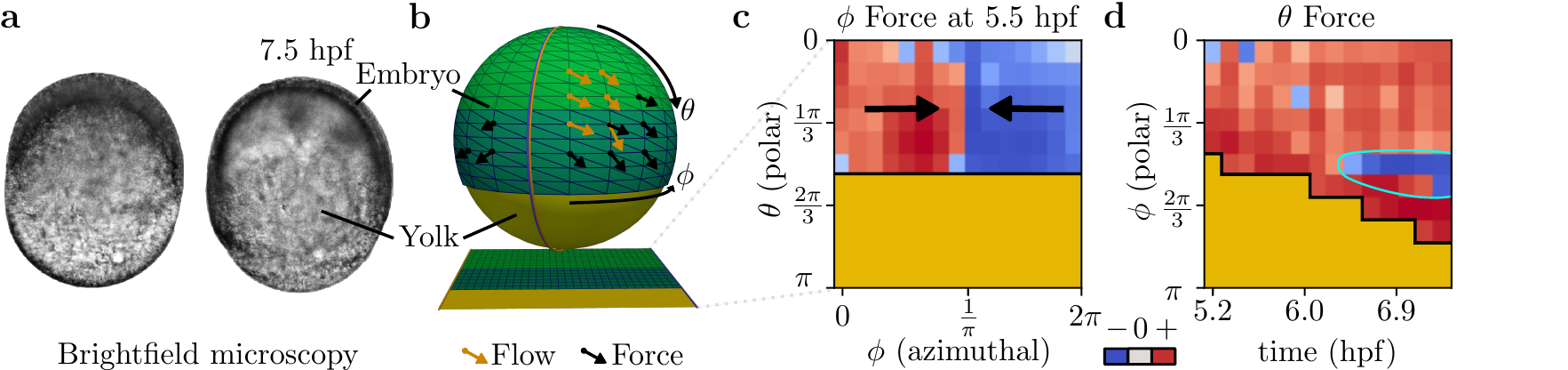}
    }
    \caption{Describing active forces in zebrafish embryogenesis. a: Brightfield images of different development stages (measured in hours past fertilization, hpf) of the embryo (dark) in the process of enveloping the yolk. b: Schematics of the model, the embryo is described on a two dimensional spherical cap. The state is given by the flow at each grid point, which is determined by the forces encapsulated in the trained parameters. c-d: colour maps of force components, red positive, blue negative, the black line marks the border between embryo and yolk (yellow). c: A snapshot of the $\phi$ component of the force at 5.5 hpf over the spherical surface of the embryo. Black arrows illustrate the force direction of red and blue regions. d: $\theta$ component of the force averaged over the azimuthal angle $\phi$ mapped against the polar position $\theta$ and time. Turquoise highlights region of negative, hence up toward the pole directed force at later times.}
    \label{fig:ZF}
\end{figure*}

\noindent\underline{State}: 2D velocity field on a spherical shell, representing the mean tissue movement of the embryo along the yolk (cf. \Figure\ref{fig:ZF}b).

\noindent\underline{EOM}: A Navier-Stokes model for the flow in curved geometry, including a field of active forces interpolated between certain positions and times.

\noindent\underline{Parameters}: The 2D active force vector at every position and time used for interpolation, in this case around 5000 parameters. Additional parameters per measurement allow the optimisation to rotate and time-shift measurements with respect to each other.

\noindent\underline{Input data}: Mean velocity field of 12 measured zebrafish embryos with 50 time points each. All fields are aligned in order to facilitate an universal stress tensor field describing all different measurements simultaneously.

\noindent\underline{Task}: (1) infer non-observable variables from observable data, (2) validate model assumptions, (3) extend existing models with new variables to better fit the data to the model assumptions.

\noindent\underline{Results}: In this experimental setup, where the data was collected by Brightfield microscopy (an example is shown in \Figure\ref{fig:ZF}a), the ground truth is unknown, hence the results are compared to well-known details of zebrafish embryogenesis. However, the force distributions, as illustrated in \Figure\ref{fig:ZF}b), obtained from the simple model are not only plausibly explaining the spreading via a strong force at the leading embryo edge (\Figure\ref{fig:ZF}d), but also suggest two known deviations from a uniform spreading: Firstly, the $\phi$ force shows a strong dipole where cells converge at azimuthal position $\phi=\pi$ (\Figure\ref{fig:ZF}c), yielding a thickening of the spreading embryo tissue known as shield formation \cite{kimmel1995stages}. Secondly, the $\theta$ distribution (\Figure\ref{fig:ZF}d) shows at later times partly negative values, in contrast to mostly positive values corresponding to a force spreading the embryo downward around the yolk. This corroborates the established invagination of cells on the inside of the spreading embryonic tissue sheet moving back towards the pole \cite{kimmel1995stages}. However, a model including the missing radial dimension is necessary to conclusively resolve the invagination.

\section{\label{sec:conclusion}Conclusions}
AdoptODE is a novel model fitting approach that requires only data and general model assumptions in the form of ODEs. The current study has shown that it can be applied to multiple and very different systems at a high level of performance.

We could reliably reconstruct system parameters as well as states. Note that parameters could be found even if their influence on the system is only perturbative (as masses in the solar system) or only present at specific time points (as the radius in collisional N-body systems, Sec.~\ref{sec:results_nb}). 

AdoptODE allows fitting of thousands of parameters, with which it becomes possible to explore systems such as zebrafish development \ref{sec:results_ze}. It allows estimating the forces occurring in the zebrafish embryo using an unrestricted grid and an interpolation-based description of the vector field, hence with fewer spatio-temporal constraints as in the existing models \cite{morita2017physical,wallmeyer2018collective}. 
 
The BOCF example \ref{sec:results_bocf} shows that adoptODE can successfully operate in systems with large numbers of coupled equations where the optimisation problem is difficult due to the heavy cross-linking of the parameters. It is also worth noting that even a system with a $\lambda_{\text{max}}\approx 0.00276$ can be reliably optimised with a very small amount of data, potentially leading to data assimilation applications such as those to be found in personalised medicine in the future, where a personalised model for the patient has to be created based on a short measurement/data collection.

The Rayleigh-B\'enard convection example was used by us to show that initial conditions can also be considered as parameters of an ODE. We have demonstrated in Sec.~\ref{sec:results_rbc} that an unobserved part of the Rayleigh-B\'enard convection (the temperature field) can be reconstructed by fitting initial conditions reproducing the observed part of the systems development.

Again, we have a chaotic system where even small errors in the initial parameters lead to an exponentially increasing error, especially when an entire initial field has to be optimised, i.e. many parameters (here $100 \times 100$ or even $>\num{5e5}$ for the AP model recovery), this is a challenging task. However, adoptODE is able to reliably optimise even such a case and produce interpretable and understandable results with little data.

Hence, the here-treated set of problems shows that adoptODE can be used to (1) infer non-observable variables from observable data, (2) validate model assumptions, (3) extend existing models with new variables to better fit the data to the model assumptions and (4) construct low-dimensional predictive models. Commonly, the complexity of the here introduced, but also other possible application cases, requires some form of regularisation, especially if one wants to work with experimental data. This is possible in adoptODE, too, because the loss function can be freely augmented by the required regularisation terms.

Overall, adoptODE provides a highly versatile and easy-to-use framework to create clear cut parameter inferences based on fundamental physical laws, even in the case of highly complex and nonlinear interactions. Compared to other methods, is more performant and uses less memory than classical approaches, a comparison can be found in the  supplement (Sec. III A).

This tool is desperately needed in our modern world, to systematically exploit the steady increase in precises data assessment and thus obtaining the optimal knowledge of hidden variables.

\section{\label{sec:method}Method}

Each optimisation step starts by solving the EOM, obtaining the trajectory dependent on the current guesses for parameters and on the initial conditions. A loss function $L$ is used to compare the found trajectory with the data, where the choice of the loss function remains free to the user (Here we used simply the Euclidean distance between observed and inferred trajectory). The adjoint method afterwards solves the EOM backwards in time, simultaneously solving for the adjoint state $a(t)$ (which measures the sensitivity of the loss function to changes in the system state $\state$ at time $t$) and the gradients $g$ depending on the loss function $L$. A derivation of the expressions and the definition of the adjoint state are provided in \Sec{\ref{sec:Theory}}. The necessity of solving the equation backwards introduces problems,  e.g. for diffusion terms encountered in many applications, as these term become highly unstable for reversed time. To mitigate this, additional checkpoints are used to reset the solution. To optimise memory usage, these checkpoints are not all saved during the forward pass, but are generated by additional forward passes as needed \cite{ma2021comparison}.

\subsection{Derivation of the adjoint method}\label{sec:Theory}
The adoptODE system is defined by a state $\state$ that is evolving according to an equation of motion (EOM) $\eom$. This EOM depends on additional parameters $\params$. Supplemented with possibly inaccurate initial conditions $\istate$ the system is given as typical ODE problem,
\begin{align}\label{eq:ODE}
    &\tder{\state} = \eomof, &&\state\at{t=0} = \istate.
\end{align}
The task is to find parameters $\params$ minimising a real-valued loss function $\loss \left( y\at{\ot_0},...,y\at{\ot_N}\right)$, which depends on the state $\state$ of the system at a finite number of ascending evaluation time points $\ot_0, \ot_1,...,\ot_N$. To be found are the parameters $\params$ and possibly initial conditions $\istate$ such that the loss is minimal. Finding this minimum is non-trivial, because the state $\state$, which enters the loss function, depends on parameters $\params$ implicitly, by being a solution of \Eq{\ref{eq:ODE}}. Furthermore, solutions at different times are connected, therefore the different $\state\at{\ot_i}$ cannot be optimised independently. This problem had been studied in detail in \cite{pontryagin1987mathematical}. To simplify notation, state $\state$ and parameters $\params$ are treated here as scalars. The generalisation to more dimensions is straightforward.

\subsection{Defining the Adjoint State}
We use two different possibilities for differentiating the solution of an ODE or functions $Q$ derived from it (e.g. $\loss$) at a time t with respect to some quantity $X$ the solution depends on. Here these quantities are the state $\state\at{t}$ and the parameters $\params$:
\begin{align}
    &\text{Momentary:} &&\momder{X(t)}{Q\left(\state\at{t_0},...\right)}\label{eq:Momentary}\\
    &\text{Lasting:} &&\conder{X(t)}{Q\left(\state\at{t_0},...\right)}\label{eq:Continued}
\end{align}
The momentary derivative at $t$ is the change in $Q$ if $X$ is varied, but only at time t, and for all future times $X$ reverses to its previous value, analogous to the functional derivative from variational calculus with the simplification of using a delta distribution in time as a test function. The lasting derivative on the other hand changes $X(t)$ and keeps the changed value for all following times. In the case of $X=\params$, this means a change in parameters which induces a slight deviation in the state $\state$, which for the momentary derivative is propagated with the old parameter value, and for the lasting with the changed value. An important quantity defined by the lasting derivative with respect to the state $\state$ is the adjoint state
\begin{equation}
    \ad(t)=\conder{\state(t)}{\loss \left( y\at{\ot_0},...,y\at{\ot_N}\right)}.
\end{equation}
It describes the reaction in loss if at $t$ the system's state is changed, with the changed state afterwards propagated to effect all later states entering $\loss$.

\subsection{Adjoint Method}\label{sec:Adjoint}
In this framework the required gradients $\grad$ to solve the optimisation problem can be understood as the lasting derivative of $\loss$ with respect to the parameters $\params$ at time $t_0$. Using an integral, we can express these gradients in terms of the easier to handle momentary derivative,
\begin{equation}\label{eq:Gradients}
    \grad = \conder{p(\ot_0)}{\loss \left( y\at{\ot_0},...\right)}=\int_{\ot_N}^{\ot_0}dt\momder{p(t)}{\loss \left( y\at{\ot_0},...\right)}
\end{equation}
The integral will later be evaluated backward in time. Conceptually, this is because we know what happens if the parameters are changed at the end point $t=\ot_N$, namely  \textit{nothing} as the system has no time to react to the change. From this known start point, the gradients can be determined by moving the time of parameter change towards the initial time $\ot_0$. The momentary derivative can be expressed by the chain rule as product of the adjoint state and a Jacobian. This reflects the fact that momentarily changing the parameters introduces a change in the system's state, which -- after the parameters are changed back -- propagates as defined for the lasting derivative. The Jacobian $\partial \eom/\partial \params = \jac{\eom}{\params}$ describes the change in state caused by the parameter change.
\begin{equation}\label{eq:MomGradToAd}
    \momder{p(t)}{\loss \left(...\right)} = \conder{\state(t)}{\loss \left(...\right)}\cdot\jacof{\eom}{\params}=\ad(t)\cdot\jacof{\eom}{\params}\
\end{equation}
Combining \Eq{\ref{eq:Gradients}} and \Eq{\ref{eq:MomGradToAd}}, we obtain an expression for the time derivative of the gradient $\grad$,
\begin{equation}\label{eq:GradientDyn}
    \tder{\grad(t)} = -\ad(t)\frac{\partial \eomof}{\partial p}.
\end{equation}
The desired gradients are $g=g(t_0)$, and the initial condition for solving this backwards is $g(t_N)=0$.\\
The missing piece is now the adjoint state, or an equation allowing us to derive it. For this we first consider one step between $t$ and $t+\epsilon$ with $\epsilon$ small, such that none of the evaluation times $\ot_i$ are passed within this step, as their influence has to be considered separately. Then the following holds:
\begin{equation}
\begin{split}
    \conder{y(t)}{\loss \left(...\right)}&=\conder{y(t+\epsilon)}{\loss \left(...\right)}\conder{y(t)}{y(t+\epsilon)}\\
    \Longrightarrow \ad(t)&=\ad(t+\epsilon)\conder{y(t)}{y(t+\epsilon)}
\end{split}
\end{equation}
Elementary algebra on the differential quotient yield the desired evolution of $\ad(t)$:
\begin{equation}\label{eq:AdjointDyn}
\begin{split}
    \tder{\ad(t)}=&\lim_{\epsilon\to 0}\frac{\ad(t+\epsilon)-\ad(t)}{\epsilon}\\
    =& \lim_{\epsilon\to 0}\frac{\ad(t+\epsilon)-\ad(t+\epsilon)\conder{\state(t)}{\state(t+\epsilon)}}{\epsilon}\\
    =& \lim_{\epsilon\to 0}\frac{\ad(t+\epsilon)\left(1-\left(1+\jacof{\eom}{\state}\epsilon\right)\right)}{\epsilon}\\
    =& -\ad(t)\cdot\jacof{\eom}{\state}
\end{split}
\end{equation}
The significance of the condition on not passing the evaluation time lies with the adjoint state $\ad(t)$. Defined as the change of the loss function while changing the solution at time $t$, it can be expanded as
\begin{equation}
\begin{split}
    \ad(t)=&\ \conder{\state(t)}{\loss \left( y\at{\ot_0},...,y\at{\ot_N}\right)}\\
    =&\ \sum_{i=0}^N\frac{\partial\loss \left( y\at{\ot_0},...,y\at{\ot_N}\right)}{\partial y\at{\ot_i}}\conder{y\at{t}}{y\at{\ot_i}}\\
    =&\ \sum_{i=0}^N\jac{\loss}{\state\at{\ot_i}}(\state,\ot_i,\params)\conder{y\at{t}}{y\at{\ot_i}}.
\end{split}
\end{equation}
Because $\conder{y\at{t}}{y\at{\ot_i}}$ is zero for $t>\ot_i$ and becomes unity at $t=\ot_i$, this discontinuity has to be manually incorporated into the $\ad(t)$ evolution, meaning $\jac{\loss}{\state\at{\ot_i}}(\state,\ot_i,\params)$ has to be added when $t$ passes $\ot_i$. Additionally, we find $\jac{\loss}{\state\at{\ot_N}}(\state,\ot_N,\params)$ as initial value $\ad(\ot_N)$.

\subsection{Augmented System}\label{sec:Augmented}
The results of \Sec{\ref{sec:Adjoint}} allow solving for the gradients $\grad$. After solving the equation for the state $y(t)$ forward, the gradients can be computed by solving first \Eq{\ref{eq:AdjointDyn}} for the adjoint state and then \Eq{\ref{eq:GradientDyn}}, both backwards. However, this requires storing the system state $\state$ and the adjoint state $\ad$ at a large number of intermediate times to have them available while solving \Eq{\ref{eq:GradientDyn}}, which is impractical due to high memory consumption. It is much more efficient (in terms of memory) to compute all three quantities simultaneously in a single backward pass of what is called the \textit{augmented system}.\\
The augmented system with state $\augstate$ and its time derivative is defined as
\begin{align}\label{eq:AugSys}
    \augstate(t) &= \begin{pmatrix} \state(t)\\ \ad(t)\\ \conder{\params(t)}{\loss\left(...\right)}\end{pmatrix},\\ \tder{\augstate}(t)&=\begin{pmatrix}\eomof\\ -\ad(t)\jacof{\eom}{\state}\\ -\ad(t)\jacof{\eom}{\params}\end{pmatrix},
\end{align}
where the initial condition follows from the aforementioned as
\begin{align}\label{eq:AugIni}
    \augstate(\ot_N) = \begin{pmatrix} \state\at{\ot_N}\\ \ad(\ot_N)=\jac{\loss}{\state\at{\ot_N}}(\state\at{\ot_N}, \ot_N, \params)\\ \jac{\loss}{\params}(\state\at{\ot_N}, \ot_N, \params)=0\end{pmatrix}.
\end{align}
This system can be solved backwards by solving the different co-dependent quantities $\state$, $\ad$ and $\loss$ simultaneously, which is possible as their respective EOMs are only dependent on each other evaluated at the current time $t$. An exception are the system states at observation times $\ot_i$, which are required to update the adjoint state at observation times. They can be saved during the forward pass, which is required anyways to obtain $\state\at{\ot_N}$. As only the system state and only at the observation times have to be saved, the augmented system significantly reduces memory consumption in comparison to saving the state and the augmented state at all the intervals required by the ODE solver. Another advantage is that at the final value of the adjoint state, $\ad(\ot_0)$, gradients are available to train the initial conditions.\\
The adjoint method in the form of the augmented system provides a powerful tool for solving the complex problem of obtaining the gradients $\grad$. The memory cost is linear in the dimension of  state $y$ and  parameters $\params$. The main computation necessary is to solve the ODE of the augmented system (\Eq{\ref{eq:AugSys}}), which between the observation times can be done by calling any standard ODE solver. Besides utilising a large base of well-developed ODE solvers in general, some allow tuning their accuracy, offering major performance increase if mediocre accuracy is sufficient and vice versa. Care has to be taken at each observation time $\ot_i$, as the update of the adjoint state is a non-smooth operation, which has to be performed manually outside of the ODE solver. Additionally, solving the system itself backward can introduce numerical problems, such as instabilities, not present in the forward pass, as is the case for diffusive systems. An easy workaround at the cost of more memory is to save additional backup states during the forward pass, as discussed in the supplement (Sec. 1).

\begin{acknowledgments}
The authors would like to thank the Deutsche Forschungsgemeinschaft (DFG, German Research Foundation) - Project-ID 454648639 - SFB 1528 for funding this work. SH would also like to thank Ulrich Parlitz and Christian Tetzlaff for their continued support and inspiring scientific discussions. LL would like to thank Ulrich Schwarz, Falko Ziebert and Oliver M. Drozdowski for their continuous support and many scientific discussions.
\end{acknowledgments}

\bibliography{apssamp}

\begin{thebibliography}{30}%
\makeatletter
\providecommand \@ifxundefined [1]{%
 \@ifx{#1\undefined}
}%
\providecommand \@ifnum [1]{%
 \ifnum #1\expandafter \@firstoftwo
 \else \expandafter \@secondoftwo
 \fi
}%
\providecommand \@ifx [1]{%
 \ifx #1\expandafter \@firstoftwo
 \else \expandafter \@secondoftwo
 \fi
}%
\providecommand \natexlab [1]{#1}%
\providecommand \enquote  [1]{``#1''}%
\providecommand \bibnamefont  [1]{#1}%
\providecommand \bibfnamefont [1]{#1}%
\providecommand \citenamefont [1]{#1}%
\providecommand \href@noop [0]{\@secondoftwo}%
\providecommand \href [0]{\begingroup \@sanitize@url \@href}%
\providecommand \@href[1]{\@@startlink{#1}\@@href}%
\providecommand \@@href[1]{\endgroup#1\@@endlink}%
\providecommand \@sanitize@url [0]{\catcode `\\12\catcode `\$12\catcode `\&12\catcode `\#12\catcode `\^12\catcode `\_12\catcode `\%12\relax}%
\providecommand \@@startlink[1]{}%
\providecommand \@@endlink[0]{}%
\providecommand \url  [0]{\begingroup\@sanitize@url \@url }%
\providecommand \@url [1]{\endgroup\@href {#1}{\urlprefix }}%
\providecommand \urlprefix  [0]{URL }%
\providecommand \Eprint [0]{\href }%
\providecommand \doibase [0]{https://doi.org/}%
\providecommand \selectlanguage [0]{\@gobble}%
\providecommand \bibinfo  [0]{\@secondoftwo}%
\providecommand \bibfield  [0]{\@secondoftwo}%
\providecommand \translation [1]{[#1]}%
\providecommand \BibitemOpen [0]{}%
\providecommand \bibitemStop [0]{}%
\providecommand \bibitemNoStop [0]{.\EOS\space}%
\providecommand \EOS [0]{\spacefactor3000\relax}%
\providecommand \BibitemShut  [1]{\csname bibitem#1\endcsname}%
\let\auto@bib@innerbib\@empty
\bibitem [{\citenamefont {Roney}\ and\ \citenamefont {Ovchinnikov}(2022)}]{alphafold2}%
  \BibitemOpen
  \bibfield  {author} {\bibinfo {author} {\bibfnamefont {J.~P.}\ \bibnamefont {Roney}}\ and\ \bibinfo {author} {\bibfnamefont {S.}~\bibnamefont {Ovchinnikov}},\ }\href {https://doi.org/10.1103/PhysRevLett.129.238101} {\bibfield  {journal} {\bibinfo  {journal} {Phys. Rev. Lett.}\ }\textbf {\bibinfo {volume} {129}},\ \bibinfo {pages} {238101} (\bibinfo {year} {2022})}\BibitemShut {NoStop}%
\bibitem [{\citenamefont {Breen}\ \emph {et~al.}(2020)\citenamefont {Breen}, \citenamefont {Foley}, \citenamefont {Boekholt},\ and\ \citenamefont {Zwart}}]{threebody}%
  \BibitemOpen
  \bibfield  {author} {\bibinfo {author} {\bibfnamefont {P.~G.}\ \bibnamefont {Breen}}, \bibinfo {author} {\bibfnamefont {C.~N.}\ \bibnamefont {Foley}}, \bibinfo {author} {\bibfnamefont {T.}~\bibnamefont {Boekholt}},\ and\ \bibinfo {author} {\bibfnamefont {S.~P.}\ \bibnamefont {Zwart}},\ }\href@noop {} {\bibfield  {journal} {\bibinfo  {journal} {Monthly Notices of the Royal Astronomical Society}\ }\textbf {\bibinfo {volume} {494}},\ \bibinfo {pages} {2465} (\bibinfo {year} {2020})}\BibitemShut {NoStop}%
\bibitem [{\citenamefont {Xie}\ and\ \citenamefont {Grossman}(2018)}]{TianEtAl2018}%
  \BibitemOpen
  \bibfield  {author} {\bibinfo {author} {\bibfnamefont {T.}~\bibnamefont {Xie}}\ and\ \bibinfo {author} {\bibfnamefont {J.~C.}\ \bibnamefont {Grossman}},\ }\href {https://doi.org/10.1103/PhysRevLett.120.145301} {\bibfield  {journal} {\bibinfo  {journal} {Phys. Rev. Lett.}\ }\textbf {\bibinfo {volume} {120}},\ \bibinfo {pages} {145301} (\bibinfo {year} {2018})}\BibitemShut {NoStop}%
\bibitem [{\citenamefont {Bradbury}\ \emph {et~al.}(2018)\citenamefont {Bradbury}, \citenamefont {Frostig}, \citenamefont {Hawkins}, \citenamefont {Johnson}, \citenamefont {Leary}, \citenamefont {Maclaurin}, \citenamefont {Necula}, \citenamefont {Paszke}, \citenamefont {Vander{P}las}, \citenamefont {Wanderman-{M}ilne},\ and\ \citenamefont {Zhang}}]{jax2018github}%
  \BibitemOpen
  \bibfield  {author} {\bibinfo {author} {\bibfnamefont {J.}~\bibnamefont {Bradbury}}, \bibinfo {author} {\bibfnamefont {R.}~\bibnamefont {Frostig}}, \bibinfo {author} {\bibfnamefont {P.}~\bibnamefont {Hawkins}}, \bibinfo {author} {\bibfnamefont {M.~J.}\ \bibnamefont {Johnson}}, \bibinfo {author} {\bibfnamefont {C.}~\bibnamefont {Leary}}, \bibinfo {author} {\bibfnamefont {D.}~\bibnamefont {Maclaurin}}, \bibinfo {author} {\bibfnamefont {G.}~\bibnamefont {Necula}}, \bibinfo {author} {\bibfnamefont {A.}~\bibnamefont {Paszke}}, \bibinfo {author} {\bibfnamefont {J.}~\bibnamefont {Vander{P}las}}, \bibinfo {author} {\bibfnamefont {S.}~\bibnamefont {Wanderman-{M}ilne}},\ and\ \bibinfo {author} {\bibfnamefont {Q.}~\bibnamefont {Zhang}},\ }\href {http://github.com/google/jax} {\bibinfo {title} {{JAX}: composable transformations of {P}ython+{N}um{P}y programs}} (\bibinfo {year} {2018})\BibitemShut {NoStop}%
\bibitem [{\citenamefont {Tjoa}\ and\ \citenamefont {Biegler}(1991)}]{tjoa1991simultaneous}%
  \BibitemOpen
  \bibfield  {author} {\bibinfo {author} {\bibfnamefont {I.~B.}\ \bibnamefont {Tjoa}}\ and\ \bibinfo {author} {\bibfnamefont {L.~T.}\ \bibnamefont {Biegler}},\ }\href@noop {} {\bibfield  {journal} {\bibinfo  {journal} {Industrial \& Engineering Chemistry Research}\ }\textbf {\bibinfo {volume} {30}},\ \bibinfo {pages} {376} (\bibinfo {year} {1991})}\BibitemShut {NoStop}%
\bibitem [{\citenamefont {Farmer}\ and\ \citenamefont {Sidorowich}(1991)}]{farmer1991optimal}%
  \BibitemOpen
  \bibfield  {author} {\bibinfo {author} {\bibfnamefont {J.~D.}\ \bibnamefont {Farmer}}\ and\ \bibinfo {author} {\bibfnamefont {J.~J.}\ \bibnamefont {Sidorowich}},\ }\href@noop {} {\bibfield  {journal} {\bibinfo  {journal} {Physica D: Nonlinear Phenomena}\ }\textbf {\bibinfo {volume} {47}},\ \bibinfo {pages} {373} (\bibinfo {year} {1991})}\BibitemShut {NoStop}%
\bibitem [{\citenamefont {Baake}\ \emph {et~al.}(1992)\citenamefont {Baake}, \citenamefont {Baake}, \citenamefont {Bock},\ and\ \citenamefont {Briggs}}]{baake1992fitting}%
  \BibitemOpen
  \bibfield  {author} {\bibinfo {author} {\bibfnamefont {E.}~\bibnamefont {Baake}}, \bibinfo {author} {\bibfnamefont {M.}~\bibnamefont {Baake}}, \bibinfo {author} {\bibfnamefont {H.}~\bibnamefont {Bock}},\ and\ \bibinfo {author} {\bibfnamefont {K.}~\bibnamefont {Briggs}},\ }\href@noop {} {\bibfield  {journal} {\bibinfo  {journal} {Physical Review A}\ }\textbf {\bibinfo {volume} {45}},\ \bibinfo {pages} {5524} (\bibinfo {year} {1992})}\BibitemShut {NoStop}%
\bibitem [{\citenamefont {Tanartkit}\ and\ \citenamefont {Biegler}(1995)}]{tanartkit1995stable}%
  \BibitemOpen
  \bibfield  {author} {\bibinfo {author} {\bibfnamefont {P.}~\bibnamefont {Tanartkit}}\ and\ \bibinfo {author} {\bibfnamefont {L.~T.}\ \bibnamefont {Biegler}},\ }\href@noop {} {\bibfield  {journal} {\bibinfo  {journal} {Industrial \& engineering chemistry research}\ }\textbf {\bibinfo {volume} {34}},\ \bibinfo {pages} {1253} (\bibinfo {year} {1995})}\BibitemShut {NoStop}%
\bibitem [{\citenamefont {Li}\ \emph {et~al.}(2005)\citenamefont {Li}, \citenamefont {Osborne},\ and\ \citenamefont {Prvan}}]{li2005parameter}%
  \BibitemOpen
  \bibfield  {author} {\bibinfo {author} {\bibfnamefont {Z.}~\bibnamefont {Li}}, \bibinfo {author} {\bibfnamefont {M.~R.}\ \bibnamefont {Osborne}},\ and\ \bibinfo {author} {\bibfnamefont {T.}~\bibnamefont {Prvan}},\ }\href@noop {} {\bibfield  {journal} {\bibinfo  {journal} {IMA Journal of Numerical Analysis}\ }\textbf {\bibinfo {volume} {25}},\ \bibinfo {pages} {264} (\bibinfo {year} {2005})}\BibitemShut {NoStop}%
\bibitem [{\citenamefont {Brewer}\ \emph {et~al.}(2008)\citenamefont {Brewer}, \citenamefont {Barenco}, \citenamefont {Callard}, \citenamefont {Hubank},\ and\ \citenamefont {Stark}}]{brewer2008fitting}%
  \BibitemOpen
  \bibfield  {author} {\bibinfo {author} {\bibfnamefont {D.}~\bibnamefont {Brewer}}, \bibinfo {author} {\bibfnamefont {M.}~\bibnamefont {Barenco}}, \bibinfo {author} {\bibfnamefont {R.}~\bibnamefont {Callard}}, \bibinfo {author} {\bibfnamefont {M.}~\bibnamefont {Hubank}},\ and\ \bibinfo {author} {\bibfnamefont {J.}~\bibnamefont {Stark}},\ }\href@noop {} {\bibfield  {journal} {\bibinfo  {journal} {Philosophical Transactions of the Royal Society A: Mathematical, Physical and Engineering Sciences}\ }\textbf {\bibinfo {volume} {366}},\ \bibinfo {pages} {519} (\bibinfo {year} {2008})}\BibitemShut {NoStop}%
\bibitem [{\citenamefont {Chen}\ \emph {et~al.}(2018)\citenamefont {Chen}, \citenamefont {Rubanova}, \citenamefont {Bettencourt},\ and\ \citenamefont {Duvenaud}}]{chen2018neural}%
  \BibitemOpen
  \bibfield  {author} {\bibinfo {author} {\bibfnamefont {R.~T.~Q.}\ \bibnamefont {Chen}}, \bibinfo {author} {\bibfnamefont {Y.}~\bibnamefont {Rubanova}}, \bibinfo {author} {\bibfnamefont {J.}~\bibnamefont {Bettencourt}},\ and\ \bibinfo {author} {\bibfnamefont {D.~K.}\ \bibnamefont {Duvenaud}},\ }in\ \href {https://proceedings.neurips.cc/paper/2018/file/69386f6bb1dfed68692a24c8686939b9-Paper.pdf} {\emph {\bibinfo {booktitle} {Advances in Neural Information Processing Systems}}},\ Vol.~\bibinfo {volume} {31},\ \bibinfo {editor} {edited by\ \bibinfo {editor} {\bibfnamefont {S.}~\bibnamefont {Bengio}}, \bibinfo {editor} {\bibfnamefont {H.}~\bibnamefont {Wallach}}, \bibinfo {editor} {\bibfnamefont {H.}~\bibnamefont {Larochelle}}, \bibinfo {editor} {\bibfnamefont {K.}~\bibnamefont {Grauman}}, \bibinfo {editor} {\bibfnamefont {N.}~\bibnamefont {Cesa-Bianchi}},\ and\ \bibinfo {editor} {\bibfnamefont {R.}~\bibnamefont {Garnett}}}\ (\bibinfo  {publisher} {Curran Associates, Inc.},\ \bibinfo {year} {2018})\BibitemShut {NoStop}%
\bibitem [{\citenamefont {He}\ \emph {et~al.}(2016)\citenamefont {He}, \citenamefont {Zhang}, \citenamefont {Ren},\ and\ \citenamefont {Sun}}]{resnet}%
  \BibitemOpen
  \bibfield  {author} {\bibinfo {author} {\bibfnamefont {K.}~\bibnamefont {He}}, \bibinfo {author} {\bibfnamefont {X.}~\bibnamefont {Zhang}}, \bibinfo {author} {\bibfnamefont {S.}~\bibnamefont {Ren}},\ and\ \bibinfo {author} {\bibfnamefont {J.}~\bibnamefont {Sun}},\ }in\ \href {https://doi.org/10.1109/CVPR.2016.90} {\emph {\bibinfo {booktitle} {2016 IEEE Conference on Computer Vision and Pattern Recognition (CVPR)}}}\ (\bibinfo {year} {2016})\ pp.\ \bibinfo {pages} {770--778}\BibitemShut {NoStop}%
\bibitem [{\citenamefont {Lotka}(1910)}]{lotka2002contribution}%
  \BibitemOpen
  \bibfield  {author} {\bibinfo {author} {\bibfnamefont {A.~J.}\ \bibnamefont {Lotka}},\ }\href@noop {} {\bibfield  {journal} {\bibinfo  {journal} {The Journal of Physical Chemistry}\ }\textbf {\bibinfo {volume} {14}},\ \bibinfo {pages} {271} (\bibinfo {year} {1910})}\BibitemShut {NoStop}%
\bibitem [{\citenamefont {Volterra}(1926)}]{volterra1926fluctuations}%
  \BibitemOpen
  \bibfield  {author} {\bibinfo {author} {\bibfnamefont {V.}~\bibnamefont {Volterra}},\ }\href@noop {} {\bibfield  {journal} {\bibinfo  {journal} {Nature}\ }\textbf {\bibinfo {volume} {118}},\ \bibinfo {pages} {558} (\bibinfo {year} {1926})}\BibitemShut {NoStop}%
\bibitem [{\citenamefont {Newton}(1833)}]{newton1833philosophiae}%
  \BibitemOpen
  \bibfield  {author} {\bibinfo {author} {\bibfnamefont {I.}~\bibnamefont {Newton}},\ }\href@noop {} {\emph {\bibinfo {title} {Philosophiae naturalis principia mathematica}}},\ Vol.~\bibinfo {volume} {1}\ (\bibinfo  {publisher} {G. Brookman},\ \bibinfo {year} {1833})\BibitemShut {NoStop}%
\bibitem [{\citenamefont {Bueno-Orovio}\ \emph {et~al.}(2008)\citenamefont {Bueno-Orovio}, \citenamefont {Cherry},\ and\ \citenamefont {Fenton}}]{bueno2008minimal}%
  \BibitemOpen
  \bibfield  {author} {\bibinfo {author} {\bibfnamefont {A.}~\bibnamefont {Bueno-Orovio}}, \bibinfo {author} {\bibfnamefont {E.~M.}\ \bibnamefont {Cherry}},\ and\ \bibinfo {author} {\bibfnamefont {F.~H.}\ \bibnamefont {Fenton}},\ }\href@noop {} {\bibfield  {journal} {\bibinfo  {journal} {Journal of theoretical biology}\ }\textbf {\bibinfo {volume} {253}},\ \bibinfo {pages} {544} (\bibinfo {year} {2008})}\BibitemShut {NoStop}%
\bibitem [{\citenamefont {Rayleigh}(1916)}]{rayleigh1916lix}%
  \BibitemOpen
  \bibfield  {author} {\bibinfo {author} {\bibfnamefont {L.}~\bibnamefont {Rayleigh}},\ }\href@noop {} {\bibfield  {journal} {\bibinfo  {journal} {The London, Edinburgh, and Dublin Philosophical Magazine and Journal of Science}\ }\textbf {\bibinfo {volume} {32}},\ \bibinfo {pages} {529} (\bibinfo {year} {1916})}\BibitemShut {NoStop}%
\bibitem [{\citenamefont {Giorgini}(2015)}]{giorgini2015status}%
  \BibitemOpen
  \bibfield  {author} {\bibinfo {author} {\bibfnamefont {J.~D.}\ \bibnamefont {Giorgini}},\ }\href@noop {} {\bibfield  {journal} {\bibinfo  {journal} {IAU General Assembly}\ }\textbf {\bibinfo {volume} {29}},\ \bibinfo {pages} {2256293} (\bibinfo {year} {2015})}\BibitemShut {NoStop}%
\bibitem [{\citenamefont {Jones}(1924)}]{jones1924determination}%
  \BibitemOpen
  \bibfield  {author} {\bibinfo {author} {\bibfnamefont {J.~E.}\ \bibnamefont {Jones}},\ }\href@noop {} {\bibfield  {journal} {\bibinfo  {journal} {Proceedings of the Royal Society of London. Series A, Containing Papers of a Mathematical and Physical Character}\ }\textbf {\bibinfo {volume} {106}},\ \bibinfo {pages} {441} (\bibinfo {year} {1924})}\BibitemShut {NoStop}%
\bibitem [{\citenamefont {Aliev}\ and\ \citenamefont {Panfilov}(1996)}]{aliev1996simple}%
  \BibitemOpen
  \bibfield  {author} {\bibinfo {author} {\bibfnamefont {R.~R.}\ \bibnamefont {Aliev}}\ and\ \bibinfo {author} {\bibfnamefont {A.~V.}\ \bibnamefont {Panfilov}},\ }\href@noop {} {\bibfield  {journal} {\bibinfo  {journal} {Chaos, Solitons \& Fractals}\ }\textbf {\bibinfo {volume} {7}},\ \bibinfo {pages} {293} (\bibinfo {year} {1996})}\BibitemShut {NoStop}%
\bibitem [{\citenamefont {Oberbeck}(1879)}]{oberbeck1879warmeleitung}%
  \BibitemOpen
  \bibfield  {author} {\bibinfo {author} {\bibfnamefont {A.}~\bibnamefont {Oberbeck}},\ }\href@noop {} {\bibfield  {journal} {\bibinfo  {journal} {Annalen der Physik}\ }\textbf {\bibinfo {volume} {243}},\ \bibinfo {pages} {271} (\bibinfo {year} {1879})}\BibitemShut {NoStop}%
\bibitem [{\citenamefont {Boussinesq}(1903)}]{boussinesq1903thorie}%
  \BibitemOpen
  \bibfield  {author} {\bibinfo {author} {\bibfnamefont {J.}~\bibnamefont {Boussinesq}},\ }\href@noop {} {\emph {\bibinfo {title} {Theorie analytique de la chaleur mise en harmonie avec la thermodynamique et avec la theorie mcanique de la lumire: Refroidissement et chauffement par rayonnement, conductibilti des tiges, lames et masses cristallines, courants de convection, theorie mcanique de la lumire. 1903. xxxii, 625,[1] p}}},\ Vol.~\bibinfo {volume} {2}\ (\bibinfo  {publisher} {Gauthier-Villars},\ \bibinfo {year} {1903})\BibitemShut {NoStop}%
\bibitem [{\citenamefont {Rempfer}(2006)}]{rempfer2006boundary}%
  \BibitemOpen
  \bibfield  {author} {\bibinfo {author} {\bibfnamefont {D.}~\bibnamefont {Rempfer}},\ }\href {https://doi.org/10.1115/1.2177683} {\bibfield  {journal} {\bibinfo  {journal} {Applied Mechanics Reviews}\ }\textbf {\bibinfo {volume} {59}},\ \bibinfo {pages} {107} (\bibinfo {year} {2006})}\BibitemShut {NoStop}%
\bibitem [{\citenamefont {Grant}(1997)}]{grant1997particle}%
  \BibitemOpen
  \bibfield  {author} {\bibinfo {author} {\bibfnamefont {I.}~\bibnamefont {Grant}},\ }\href@noop {} {\bibfield  {journal} {\bibinfo  {journal} {Proceedings of the Institution of Mechanical Engineers, Part C: Journal of Mechanical Engineering Science}\ }\textbf {\bibinfo {volume} {211}},\ \bibinfo {pages} {55} (\bibinfo {year} {1997})}\BibitemShut {NoStop}%
\bibitem [{\citenamefont {Amat}\ \emph {et~al.}(2014)\citenamefont {Amat}, \citenamefont {Lemon}, \citenamefont {Mossing}, \citenamefont {McDole}, \citenamefont {Wan}, \citenamefont {Branson}, \citenamefont {Myers},\ and\ \citenamefont {Keller}}]{amat2014fast}%
  \BibitemOpen
  \bibfield  {author} {\bibinfo {author} {\bibfnamefont {F.}~\bibnamefont {Amat}}, \bibinfo {author} {\bibfnamefont {W.}~\bibnamefont {Lemon}}, \bibinfo {author} {\bibfnamefont {D.~P.}\ \bibnamefont {Mossing}}, \bibinfo {author} {\bibfnamefont {K.}~\bibnamefont {McDole}}, \bibinfo {author} {\bibfnamefont {Y.}~\bibnamefont {Wan}}, \bibinfo {author} {\bibfnamefont {K.}~\bibnamefont {Branson}}, \bibinfo {author} {\bibfnamefont {E.~W.}\ \bibnamefont {Myers}},\ and\ \bibinfo {author} {\bibfnamefont {P.~J.}\ \bibnamefont {Keller}},\ }\href@noop {} {\bibfield  {journal} {\bibinfo  {journal} {Nature methods}\ }\textbf {\bibinfo {volume} {11}},\ \bibinfo {pages} {951} (\bibinfo {year} {2014})}\BibitemShut {NoStop}%
\bibitem [{\citenamefont {Kimmel}\ \emph {et~al.}(1995)\citenamefont {Kimmel}, \citenamefont {Ballard}, \citenamefont {Kimmel}, \citenamefont {Ullmann},\ and\ \citenamefont {Schilling}}]{kimmel1995stages}%
  \BibitemOpen
  \bibfield  {author} {\bibinfo {author} {\bibfnamefont {C.~B.}\ \bibnamefont {Kimmel}}, \bibinfo {author} {\bibfnamefont {W.~W.}\ \bibnamefont {Ballard}}, \bibinfo {author} {\bibfnamefont {S.~R.}\ \bibnamefont {Kimmel}}, \bibinfo {author} {\bibfnamefont {B.}~\bibnamefont {Ullmann}},\ and\ \bibinfo {author} {\bibfnamefont {T.~F.}\ \bibnamefont {Schilling}},\ }\href@noop {} {\bibfield  {journal} {\bibinfo  {journal} {Developmental dynamics}\ }\textbf {\bibinfo {volume} {203}},\ \bibinfo {pages} {253} (\bibinfo {year} {1995})}\BibitemShut {NoStop}%
\bibitem [{\citenamefont {Morita}\ \emph {et~al.}(2017)\citenamefont {Morita}, \citenamefont {Grigolon}, \citenamefont {Bock}, \citenamefont {Krens}, \citenamefont {Salbreux},\ and\ \citenamefont {Heisenberg}}]{morita2017physical}%
  \BibitemOpen
  \bibfield  {author} {\bibinfo {author} {\bibfnamefont {H.}~\bibnamefont {Morita}}, \bibinfo {author} {\bibfnamefont {S.}~\bibnamefont {Grigolon}}, \bibinfo {author} {\bibfnamefont {M.}~\bibnamefont {Bock}}, \bibinfo {author} {\bibfnamefont {S.~G.}\ \bibnamefont {Krens}}, \bibinfo {author} {\bibfnamefont {G.}~\bibnamefont {Salbreux}},\ and\ \bibinfo {author} {\bibfnamefont {C.-P.}\ \bibnamefont {Heisenberg}},\ }\href@noop {} {\bibfield  {journal} {\bibinfo  {journal} {Developmental cell}\ }\textbf {\bibinfo {volume} {40}},\ \bibinfo {pages} {354} (\bibinfo {year} {2017})}\BibitemShut {NoStop}%
\bibitem [{\citenamefont {Wallmeyer}\ \emph {et~al.}(2018)\citenamefont {Wallmeyer}, \citenamefont {Trinschek}, \citenamefont {Yigit}, \citenamefont {Thiele},\ and\ \citenamefont {Betz}}]{wallmeyer2018collective}%
  \BibitemOpen
  \bibfield  {author} {\bibinfo {author} {\bibfnamefont {B.}~\bibnamefont {Wallmeyer}}, \bibinfo {author} {\bibfnamefont {S.}~\bibnamefont {Trinschek}}, \bibinfo {author} {\bibfnamefont {S.}~\bibnamefont {Yigit}}, \bibinfo {author} {\bibfnamefont {U.}~\bibnamefont {Thiele}},\ and\ \bibinfo {author} {\bibfnamefont {T.}~\bibnamefont {Betz}},\ }\href@noop {} {\bibfield  {journal} {\bibinfo  {journal} {Biophysical Journal}\ }\textbf {\bibinfo {volume} {114}},\ \bibinfo {pages} {213} (\bibinfo {year} {2018})}\BibitemShut {NoStop}%
\bibitem [{\citenamefont {Ma}\ \emph {et~al.}(2021)\citenamefont {Ma}, \citenamefont {Dixit}, \citenamefont {Innes}, \citenamefont {Guo},\ and\ \citenamefont {Rackauckas}}]{ma2021comparison}%
  \BibitemOpen
  \bibfield  {author} {\bibinfo {author} {\bibfnamefont {Y.}~\bibnamefont {Ma}}, \bibinfo {author} {\bibfnamefont {V.}~\bibnamefont {Dixit}}, \bibinfo {author} {\bibfnamefont {M.~J.}\ \bibnamefont {Innes}}, \bibinfo {author} {\bibfnamefont {X.}~\bibnamefont {Guo}},\ and\ \bibinfo {author} {\bibfnamefont {C.}~\bibnamefont {Rackauckas}},\ }in\ \href@noop {} {\emph {\bibinfo {booktitle} {2021 IEEE High Performance Extreme Computing Conference (HPEC)}}}\ (\bibinfo {organization} {IEEE},\ \bibinfo {year} {2021})\ pp.\ \bibinfo {pages} {1--9}\BibitemShut {NoStop}%
\bibitem [{\citenamefont {Pontryagin}(1987)}]{pontryagin1987mathematical}%
  \BibitemOpen
  \bibfield  {author} {\bibinfo {author} {\bibfnamefont {L.~S.}\ \bibnamefont {Pontryagin}},\ }\href@noop {} {\emph {\bibinfo {title} {Mathematical theory of optimal processes}}}\ (\bibinfo  {publisher} {CRC press},\ \bibinfo {year} {1987})\BibitemShut {NoStop}%
\end{thebibliography}%

\end{document}



\newcommand{\Eq}[1]{Eq. #1}
\newcommand{\Sec}[1]{Section #1}
\newcommand{\Fig}[1]{Fig. #1}
\newcommand{\Figure}{Fig.~}
\newcommand{\code}[1]{\texttt{#1}}
\newcommand{\kw}[1]{`#1`}

\newcommand{\ot}{\hat{t}}
\newcommand{\state}{y}
\newcommand{\istate}{\state_0}
\newcommand{\augstate}{s}
\newcommand{\eom}{f}
\newcommand{\params}{p}
\newcommand{\loss}{L}
\newcommand{\ad}{a}
\newcommand{\grad}{g}
\newcommand{\jac}[2]{\mathcal{J}_{#1,#2}}

\newcommand{\eomof}{f(\state,t,\params)}
\newcommand{\jacof}[2]{\mathcal{J}_{#1,#2}(\state, t, \params)}

\newcommand{\tder}[1]{\frac{d}{dt}#1}
\newcommand{\momder}[2]{\frac{\partial #2}{\partial #1}}
\newcommand{\conder}[2]{\frac{\mathcal{D} #2}{\mathcal{D} #1}}

\newcommand{\at}[1]{_{(#1)}}

\preprint{APS/123-QED}

\title{physNODE: Fusion of data and expert knowledge for modeling dynamical systems Supplements}

\author{Leon Lettermann}
 \author{Alejandro Jurado}
 \author{Timo Betz}
 \author{Florentin  W\"org\"otter}
\author{Sebastian Herzog}%
 \email{sherzog3@gwdg.de}
\affiliation{%
 Third Institute of Physics - Biophysics, Georg-August Universit\"at G\"ottingen, 
                 G\"ottingen, Deutschland 37077\\
}%




\date{\today}

\maketitle


\section{\label{sec:Implementation} Implementation Details of physNODE}
PhysNODE aims to make the adjoint method easily accessible for researchers who want to find parameters describing some ODE system. Therefore, the goal is to provide a software tool that requires only minimal information to be entered, essentially the EOM of the system under consideration.  Consequently, the Jacobians in the augmented system \Eq{11,12} have to be derived internally and, additionally, this step needs to be very performant, as it will be called many times while solving the augmented system. Further aspects include possibilities to run simulations in order to validate the fitting accuracy and fitting multiple experiments with a single model.\\

An important ingredient for achieving the aforementioned goals is the use of JAX \cite{jax2018github} as a computational framework. JAX is a python library developed for machine learning applications as an open-source project with support from Google Research. It is built for just-in-time compiling its functions using an Accelerated Linear Algebra (XLA) compiler, a compiler for efficient computation with backends for GPUs and TPUs developed to speed up TensorFlow code, but JAX is built more consequently than TensorFlow to suit and utilize XLA's capabilities. In particular, JAX employs a pure function formalism and is described as a set of composable transformations of functions, including
\begin{itemize}
    \item \code{jit}: just-in-time compilation,
    \item \code{grad}: autodifferentiation,
    \item \code{vmap}: vectorization.
\end{itemize}

Combination of \code{grad} (or the lower level \code{vjp}) and \code{jit} applied to the EOM yield the performant Jacobians we need, while the \code{vmap} transformation is used to support multi-experiment fitting. The restriction that the provided EOM has to be compatible with JAX is mitigated by the JAX.numpy module, which rebuilds the popular and well-documented NumPy API. A further advantage is the concept of PyTrees, nested combinations of lists, tuples, dictionaries, and arrays, which are used as type for parameters and system states, providing freedom to organise these. Finally, JAX automatically detects possibly present GPUs or even TPUs, and the code will be compiled and executed using the best available option. \\

Apart from the data, the input has to be provided as a \code{define\_system} function, the details of which are given in the physNODE Cookbook (see Fig.~\ref{fig:cookbook}). Passed a number of keywords (\code{kwargs\_sys}) defining system properties, it returns four functions:
\begin{itemize}
    \item \code{gen\_y0()}: Generating an initial state.
    \item \code{gen\_params()}: Generating a set of parameters.
    \item \code{eom}: The systems Equation of motion, above $\eom$.
    \item \code{loss}: The loss function, above $\loss$.
\end{itemize}
The state and parameters returned by the first two functions can be any PyTree, usually a dictionary is clearest. \\

The physNODE module includes the \code{EquationNODE} class, wrapping the generation of the augmented system \Eq{11,12}, the \code{PhysNodeDataset} class to combine this equations with data and the \code{simple\_simulation} function, which automatically simulated data and sets up a \code{PhysNodeDataset} object.\\

The \code{train\_physnode} function takes the gradients delivered by the augmented system to perform training using the well-established ADAM optimizer \cite{adam}. The hyperparameters of this optimizer are a learning rate, a learning rate decay controlling exponential decay of the learning rate as well as the ADAM internal constants $b_1$ and $b_2$, all of which can be specified via \code{kwargs\_NODE}. Other optimizers can be submitted via respective \code{kwargs\_NODE}. For each of \code{params}, \code{iparams} (cf. \Sec\ref{sec:WorkWithP}) and possibly \code{y}$_0$ that should be fitted a separate ADAM optimizer is used, the hyperparameters of which can independently be adjusted. The augmented method is executed neglecting this dependence, only at the end of each epoch are the gradient contributions generated by the explicit derivative of the loss with respect to the parameters added. \\

In \Sec{C.} the idea to solve the system itself backward alongside adjoint state and computed gradient to avoid the necessity to save many intermediate states was introduced. In practice, time asymmetric systems may thwart this idea. One such bad, but unfortunately common, example are diffusion systems. While quite helpful in stabilizing the forward pass, solving a diffusive system backward is equivalent to anti-diffusion, enlarging initial inaccuracies exponentially and thereby rendering the backward pass highly unstable.
To handle this, physNODEs is implemented in such a way that the system state is stored at each observation time $\ot_i$, allowing backpropagation between observation times without becoming too unstable. Further stabilization can be achieved by storing additional back up states  in between two observation times, the number of which is specified by the keyword \kw{N\_backups}. In order not to save all backup states simultaneously, they are created by an additional forward pass only once their respective time interval is concerned.\\

Most often the loss function will work such that the current solution is compared to a given solution, e.g. with a mean squared error loss, and the direction of change for optimization enters the augmented system \Eq{11,12} via the Jacobian $\jac{\loss}{\state}$. 

\section{\label{sec:Remarkds} Remarks and Specifics}

Two potential problems currently exist: 
\begin{enumerate}
    \item There is currently no easy way to install JAX on non-Linux systems, and even there it does not have a built-in CUDA installation, as e.g. PyTorch or Tensorflow, meaning CUDA has to be installed manually. For this reason, we provide a Apptainer recipe to set up a container with CUDA and JAX installed, see \url{https://gitlab.gwdg.de/sherzog3/physnode.git}.
    Alternatively, one have to install JAX and its dependencies manually: The installation instruction for JAX and dependencies can be found at \url{ https://github.com/google/jax#installation}

    \item Also, the standard ODE solver used is a mixed fourth/fifth-order Runge-Kutta algorithm with Dormand Prince stepsize adaption. Other solvers can be used, but have to be compatible with JAX.
\end{enumerate}

\subsection{Working with physNODE}\label{sec:WorkWithP}
In the following we would like to provide a few tips that might be helpful when implementing own cases with physNODE: \\

The shapes of the PyTrees returned  by \code{gen\_params()} and \code{gen\_y0()} have to match whatever \code{loss} and \code{eom} expect, meaningful values are only required if simulations should be generated. The \code{gen\_params()} function actually has to return three PyTrees, for three different types of parameters called \code{params}, \code{iparams} and \code{exparams}. The first two are the unknown parameters that should be fitted and differ only in multi-experiment situations, where \code{params} are universal parameters and \code{iparams} are fitted individually for each experiment. The external parameters \code{exparams} are assumed to be known, but with individual values specified for each submitted experiment, hence they cannot be built into the system's definition (which is universal for all experiments). For more complicated systems, \code{define\_system} may have subroutines describing parts or steps of the EOM. Those can be returned in an otherwise empty dictionary as a fifth return value, such that they can be accessed later. \\

Usually \code{EquationNODE} should not be called directly, but is automatically included in \code{PhysNodeDataset}. To setup a \code{PhysNodeDataset}, the \code{define\_system} function and dedicated \code{kwargs\_sys}, as well as data to be fitted and \code{t\_evals}, an array containing the times at which the data was observed, have to be passed. Lastly, a second keyword dictionary, \code{kwargs\_NODE}, contains the settings for the NODE and optimization process. Mandatory are the learning rate (\kw{lr}) and number of epochs (\kw{epochs}), but a number of properties and behaviours can be controlled through other keywords. The \code{simple\_simulation} function requires the same information, apart from data, which is generated by randomly taken initial conditions and parameters according to \code{gen\_y0} and \code{gen\_params}. Once a dataset has been generated either with data or a simulation, training can be performed by calling the \code{train\_physnode} function with the dataset as argument.\\

Often one might want to further restrict or influence the training process. One possibility is to specify boundaries, e.g. to avoid the system reaching unstable parameter values. Another popular choice is weight decay. To include it, in contrast to the above derivation we allow parameter-dependent term in the loss function $\loss \left( y\at{\ot_0},...,y\at{\ot_N}, \params\right)$. \\

A typical weight decay would now be included by adding a $L_2$-norm of the parameters to the loss function, but it allows for arbitrary complex terms guiding the training to follow specific requirements.\\

For long time series or very inaccurate initial parameters, eventually, the current solution will have diverged so far from the target trajectory, that this comparison is meaningless. This can especially be expected to happen faster for chaotic systems. A very easy solution is only to use a fraction of the data, cutting the time evolution to a sufficiently stable regime.\footnote{\code{kwargs\_NODE} keyword: \kw{t\_stop\_idx}} A second possibility is to specify an exponential decay of gradients and the adjoint state applied during backward solving, such that contributions at later times have lesser influence.\footnote{\code{kwargs\_NODE} keyword: \kw{time\_decay}} The most powerful approach is to divide the time series into many short segments and reload a new initial condition at the beginning of each segment disregarding at what state the previous segment ended up. To conveniently use this, the \kw{t\_reset\_idcs} keyword in \code{kwargs\_NODE} can be passed a tuple of integers, which are understood as indices of observation times in \code{t\_evals}. At each indicated time the system is reset to the estimated state, taken from the target trajectory. The initial state is enlarged by an additional axis running overall initial conditions at different times, such that if training of initial conditions is active all different initial conditions are trained simultaneously.\\
The \code{simple\_simulation} function is helpful for easy testing, playing around, and having fun, but serves an important purpose in using physNODE to gain information of data. As with any tool from the realm of machine and deep learning, interpretation of the results needs care and vigilance. In this setting, the question to be answered is if found parameters, which yield a good fit of observed and reproduced system behaviour, are necessarily the true parameters. The main purpose of the simulation feature is to generate several systems with plausible trajectories. After the same fitting procedure as for actual data is applied, the found parameters can be compared to those of the simulation to validate the method. The other way round simulations can also be used before fitting data to establish a protocol and hyperparameters. If parameters do not agree even though the final loss was very small, probably some parameters are redundant allowing for different parameters describing the same dynamics, or the loss function doesn't capture all relevant parts of the system.
\section{\label{sec:VidCapts}Supplementary Figures and Videos}
\subsection{Figure SF1: Initial conditions for AP Model}
\begin{figure*}
    \centering
    \includegraphics{Supplements/S1_AP_Initial.png}
    \caption{a)-d): The fixed initial conditions used for generating the BOCF trajectories, $u$ (a) being the excitation field and $v$,$w$, and $s$ (b-d) the different gating variables. e)-g): The resulting optimized initial conditions for the AP model in order to reproduce the BOCF model. Excitation field $u$ (e) and the single gating field $v$, displayed once (f) with the common color scheme used for a)-f) and once with a separate one to show the full range (g).}
    \label{fig:my_label}
\end{figure*}
\subsection{Video V1: BOCF and AP Models}
Results of BOCF model. a-d: Excitation field $u$, value given by colour bar on the right. a: Recovered BOCF field after \SI{1.5}{\second}. b: Difference between recovery in a and target. c: Truth after \SI{0.5}{\second}. d: Recovery using the Aliev-Panfilov model. e: Mean absolute error of recovery using BOCF and AP models. f: Error in peak occurrence time averaged over all pixels.
\subsection{Video V2: Rayleigh B\'enard Convection}
Results for 2D Rayleigh-B\'enard Convection. a-c: Temperature perturbation. a: Truth, b: Recovery, c: Difference. d: Mean absolute error between recovered and true temperature perturbation. e: Pixelvalue of center pixel.
\newpage
\begin{widetext}
\begin{figure}
    \label{fig:cookbook}
    \noindent\makebox[\textwidth]{
    \includegraphics[width=1.2\textwidth]{Supplements/physNODE_Cookbook.pdf}}
\end{figure}
\end{widetext}
\newpage
\bibliography{apssamp}